\documentclass[
    sigconf,
]{acmart}

\usepackage{cleveref}
\usepackage[inline]{enumitem}

\AtBeginDocument{%
  \providecommand\BibTeX{{%
    \normalfont B\kern-0.5em{\scshape i\kern-0.25em b}\kern-0.8em\TeX}}}

\copyrightyear{2024}
\acmYear{2024}
\setcopyright{acmlicensed}\acmConference[IDE '24]{2024 First IDE Workshop}{April 20, 2024}{Lisbon, Portugal}
\acmBooktitle{2024 First IDE Workshop (IDE '24), April 20, 2024, Lisbon, Portugal}
\acmDOI{10.1145/3643796.3648458}
\acmISBN{979-8-4007-0580-9/24/04}

\begin{document}

\title{HyLiMo: A Hybrid Live-Synchronized Modular Diagramming Editor as IDE Extension for Technical and Scientific Publications}

\author{Niklas Krieger}
\email{niklas.krieger@iste.uni-stuttgart.de}
\orcid{0009-0007-7616-3155}
\affiliation{%
  \institution{Institute of Software Engineering}
  \city{Stuttgart}
  \postcode{70569}
  \country{Germany}
}

\author{Sandro Speth}
\email{sandro.speth@iste.uni-stuttgart.de}
\orcid{0000-0002-9790-3702}
\affiliation{%
  \institution{Institute of Software Engineering}
  \city{Stuttgart}
  \postcode{70569}
  \country{Germany}
}

\author{Steffen Becker}
\email{steffen.becker@iste.uni-stuttgart.de}
\orcid{0000-0002-4532-1460}
\affiliation{%
  \institution{Institute of Software Engineering}
  \city{Stuttgart}
  \postcode{70569}
  \country{Germany}
}

\renewcommand{\shortauthors}{N. Krieger et al.}

\begin{abstract}
Creating suitable diagrams for technical and scientific publications is challenging and time-consuming, as manual control over the layout is required to communicate information effectively.
Existing diagramming tools usually allow modeling the diagrams via a textual domain-specific language (DSL) that can be rendered and auto-layouted or via a graphical editor.
While auto-layout is fast, the results are often not satisfying for most publications.
However, graphical editors are time-consuming to create large diagrams.
The blended or hybrid modeling concept enables creating diagrams efficiently using a DSL and editing the rendered diagram via the graphical editor for fine-tuning.
However, hybrid modeling editors are limited to individual diagram types and do not save the layout and style information in the textual description.
Therefore, we propose \textit{HyLiMo}, a hybrid live-synchronized modular diagramming editor.
In HyLiMo, diagrams are created using an internal DSL and live synchronized with an interactive graphical editor for the rendered diagram, allowing a straightforward layout and style change, which is stored in the DSL code.
HyLiMo is independent of specific diagram types, but we offer specific functionality for UML class diagrams.
Using the language server protocol, we implement it as a web app and IDE extension.
The results of our user study indicate that such an approach enables fast and precise diagramming.
\end{abstract}




\maketitle

\section{Introduction}
Diagrams such as UML class diagrams are central to information transfer for many publications, e.g., tech documentation, blog articles, and research papers.
In such use cases, the diagrams often require a precise layout to be easily understandable and space efficient.
However, creating suitable diagrams is challenging and time-consuming as it requires a significant degree of precision for creating the layout and a tool that offers all the required diagram features.
While various diagramming tools exist, they often come with their pitfalls, e.g., a lack of standard elements, especially for elaborate diagrams like UML class diagrams.
Furthermore, two approaches exist for creating diagrams, which have their unique benefits. 
Graphical editors, e.g., Visio or diagrams.net (draw.io), allow the user to create the diagrams using a graphical syntax and specify the layout manually.
This enables creating precise custom layouts but is time-consuming for complex and large diagrams.
Besides, adding or removing elements might require additional manual layouting. 
In textual editors like PlantUML or Mermaid, however, diagrams are specified textually and then automatically rendered by the editor, using an auto-layout algorithm.
This enables the creation of large and complex diagrams quickly, but creating a custom layouting is impossible.
Blended or hybrid textual-graphical modeling enables the user to use multiple notations to efficiently create and manipulate the underlying model.
However, in comparison to diagramming tools, modeling tools are more rigid as the model must conform to its metamodel.
Also, they often do not enable custom styling and adding additional elements that users might require for their publication.

Therefore, We propose \textit{HyLiMo}, a hybrid live-synchronized modular diagramming editor.
In HyLiMo, diagrams are modeled in a textual DSL and live synchronized with an interactive graphical editor for the rendered diagram, allowing a straightforward layout and style change.
Any change of the layout and style is stored in the textual description and can be put into version control systems.
Using the language server protocol (LSP), we implement HyLiMo as a standalone web app and integrate it as an IDE extension for Visual Studio Code (VS Code).
We evaluate our approach by conducting a use case study with two use cases to diagram.
The results indicate that HyLiMo enables fast and precise diagramming for publications. 

The remainder of this paper is structured as follows.
In \Cref{sec:requirements}, we briefly outline our requirements engineering process and main requirements.
\Cref{sec:hybrid-editor} elaborates on our hybrid editor and graphical diagramming.
Afterwards, we describe HyLiMo's internal DSL and underlying general-purpose programming language SyncScript in \Cref{sec:dsl-syncscript}.
Then, we explain the IDE integration in \Cref{sec:ide-integration}.
Finally, \Cref{sec:evaluation} presents our evaluation, we discuss related work in \Cref{sec:related-work} and conclude in \Cref{sec:conclusion}.

\section{Requirements Engineering}\label{sec:requirements}
To develop our concept, we conducted a requirements engineering process.
We first identified initial requirements based on the problem statement, related work, and our needs.
Next, we conducted qualitative interviews with 14 researchers in the field of computer science.
We asked the participants about their requirements, how they would imagine a textual syntax, i.e., DSL, and how they would like to use a hybrid diagramming tool, especially focusing on UML class diagrams first.
We further asked the participants to rank the elicited requirements for prioritization.
The details of our requirements engineering process and the resulting requirements are described in~\cite{krieger2023hylimo}.
Due to space reasons, we only state the most critical requirements here: (i)~styling and layout information should be stored in the DSL code using a declarative approach and live updated when editing the diagram via the graphical editor; (ii)~the DSL code should allow general-purpose language features for faster layout creation; (iii)~the DSL code should be the single source of truth; (iv)~web browser and IDE support; (v)~the DSL is extensible for additional diagram types; (vi)~different modes to layout lines should be supported; (vii)~the editor should support all UML class diagram elements; (viii)~the editor should allow auto-completion; (ix)~navigation from a diagram element in the graphical editor to the DSL code should be possible. 

\section{Hybrid Editor}\label{sec:hybrid-editor}

To enable a user-friendly blended or hybrid diagramming experience, we developed a hybrid editor.
The editor offers a textual editor supporting LSP to create diagrams using an internal DSL and a graphical editor, based on Eclipse Sprotty, to manipulate the diagrams regarding layout and styling.
Most tasks are done in the language server, which connects via the language client using LSP to the textual and graphical editor. 
Changes in the DSL code or via the graphical editor are live synchronized between both editors to provide fast feedback to the diagrammer.
We thereby chose the DSL code as the single source of truth containing the diagram elements, styling, and layout information in a textual representation.
This also allows putting the diagram code into version control systems and allows for easy integration into collaborative tools in the future.
There are two live cycles.
For a textual update, we execute the code and then re-render the diagram based on the result.
A graphical update is an interaction consisting of multiple events caused by the user interacting with the graphical view over a short period of time.
As the DSL is the diagram's single source of truth, we must update the code based on the interaction, to provide live feedback, multiple times during an interaction. 
This code update causes, as defined before, a re-execution of the code, thus updating the diagram and showing the interaction's effect.
To prevent feedback loops, we compute how the textual definition needs to be updated at the beginning of the interaction. 
For example, when starting to move a class entity, we compute how any $(x,y)$ translation will affect the textual definition and update the textual source based on this over the course of the interaction.

In our first prototype, we experienced performance issues as the execution time for the synchronization caused notable delays in particular for larger diagrams.
To mitigate these performance issues, we combine three solutions: (1)~rate-limiting interaction events, (2)~multiple language servers, and (3)~adding predictions.
(1)~For the rate limits, instead of synchronizing every edit, especially during a move action, we can wait until the last update is finished not to fill up the language server's queue for every small update.
(2)~We add multiple language servers to offload long-taking tasks, particularly rendering the diagram, to another language server and, thus, another thread to not block the language server from working on other tasks.
The language servers are implemented in a main/worker architecture by connecting the additional language servers using an LSP extension.
Execution tasks are sent to the least busy worker, and the main language server is not blocked until the execution is finished.
As large parts of the rendering process use blocking code, relying on asynchronous request handling of the language server proved insufficient.
In particular, predictions did not function as they need to be executed concurrently to the diagram rendering task.
\begin{figure}[b]
    \centering
    \includegraphics[width=1.0\linewidth]{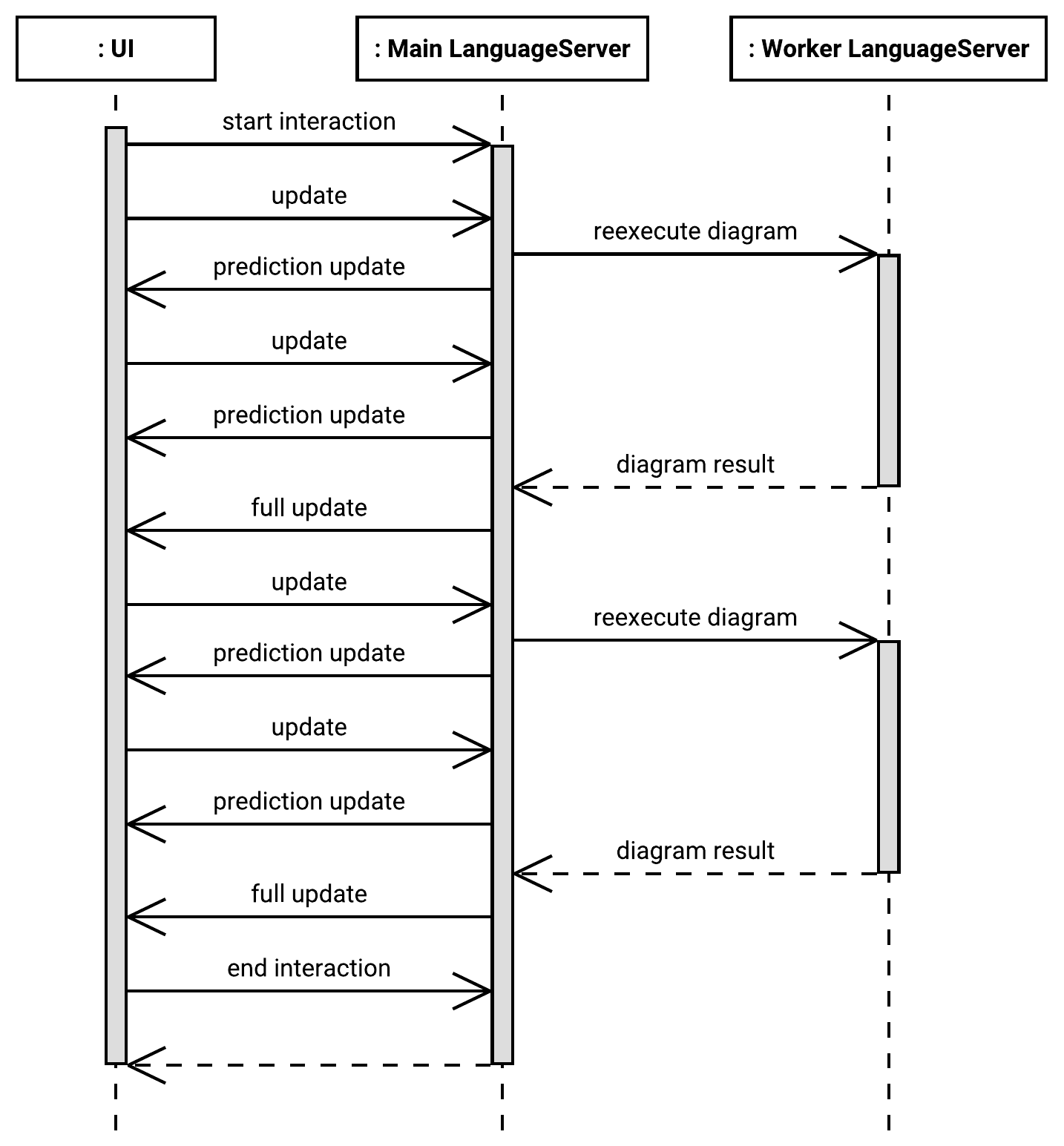}
    \caption{Sequence diagram showcasing the update \& prediction functionality.}
    \label{fig:sequence-diagram}
\end{figure}
(3)~During the interaction, we predict how the diagram is expected to change and apply this update to the diagram before the re-execution and rendering are complete.
In detail, the algorithm works as shown in \Cref{fig:sequence-diagram}:
Each type of graphical interaction, e.g., moving a diagram element, has specified parameters, such as the relative (x,y) translation.
At the start of the interaction, the server determines how, for arbitrary values for these parameters, the source code needs to be modified.
During the interaction, for each received update event, the server first updates the textual definition.
Next, the prediction is computed:
Based on the current diagram, the update used for rendering said diagram, and the current update, the delta between the two updates is computed and sent as an incremental update to the graphical view.
Last, if no full re-execution is running, a new full re-execution is started.
Yet, when the result is available, it might represent some predictions of later updates that have already happened.
Thus, the delta between the update used for the re-execution and the most current update is applied as a prediction to the diagram.
Last, the result is set as the current diagram and sent to the graphical view as a full update.
As predictions are significantly faster than full re-execution, multiple updates can be applied during one full re-execution this way, reducing input lag.
However, in some edge cases where predictions cannot be applied, notable delays cannot be avoided.
Furthermore, in some cases, predictions produce the wrong result.
This is usually caused by conditional code execution, depending on the updated layout values.
As a result, users notice flickering during the graphical interaction, caused by the incorrect prediction being overwritten regularly by the full update.
Yet, while such examples are easy to construct, such cases are rare in practice, as conditional code execution that is based on layout information is rarely used.

\section{Internal DSL and SyncScript}\label{sec:dsl-syncscript}

\begin{figure}[b]
    \centering
    \includegraphics[width=1.0\linewidth]{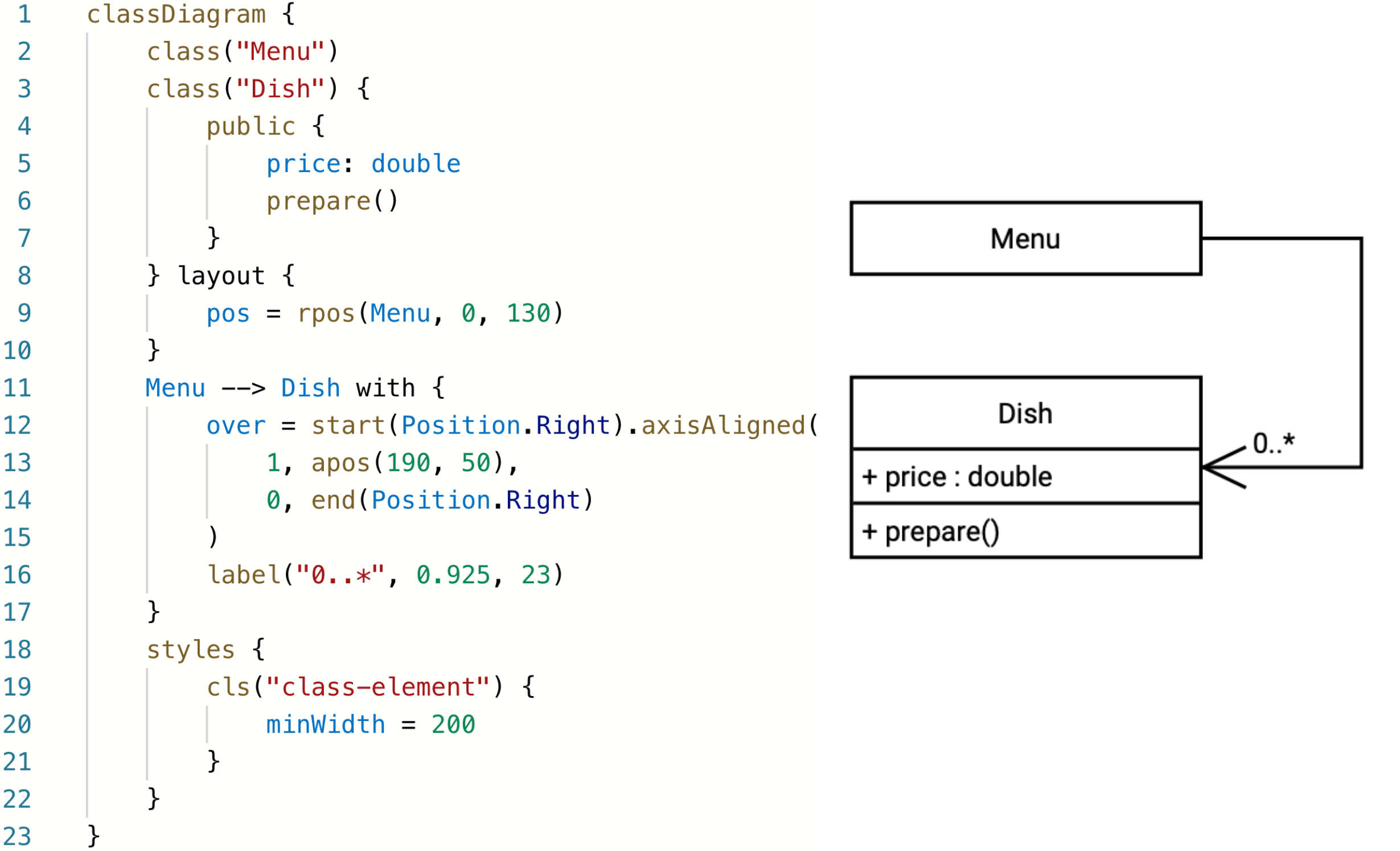}
    \caption{Example UML class diagram as textual (left) and graphical (right) view.}
    \label{fig:dsl-example}
\end{figure}

To achieve a good user experience, we require a DSL that allows as flexible as possible diagramming with syntactic sugar.
Overall, we considered two main approaches: an external DSL or an internal DSL.
External DSLs typically use a compiler-like infrastructure similar to traditional programming languages~\cite{wkasowski2023domain}.
As they are independent of other computer languages, they provide a vast flexibility in syntactic and semantics.
In contrast, internal DSLs are based on an existing general-purpose programming language (GPL).
Benefits of this approach include the availability of functionality and language features of the host GPL and the reuse of tooling for the host GPL, including editor support.
One of our main goals is to provide flexibility and extensibility to the diagrammer.
We achieve this by including GPL features in our DSL, particularly functional decomposition and imperative programming concepts.
This motivates the usage of an internal DSL, with the host language providing these features.
Hence, to build our internal DSL, we need a GPL that fulfills all our elicited requirements.
First, it must be usable in a web environment without requiring server tools and relatively fast to compile and execute, as we need to re-execute it after each edit for the live synchronization.
Further, the GPL must provide syntactic flexibility to implement the internal DSL according to our needs.
Thus, none of the considered candidates (JavaScript, Kotlin, Scala, etc.) fulfill these requirements.
While, in particular, Kotlin and Scala provide the required syntactic flexibility, execution in a browser-only environment with the required performance proved to be unfeasible.
Therefore, we decided to implemented our custom GPL, called SyncScript.
As our usage in a hybrid editor requires automatic source code modifications, designing a simple yet flexible grammar was our main design goal.
Details of SyncScript and the internal DSL can be found in~\cite{krieger2023hylimo}.
In our internal DSL design, we reuse features of other internal DSLs.
For example, we support infix operators and use lambda expressions after functions to create individual scopes.

Our internal DSL consists of multiple modules organized in three layers.
Layer 1 provides common functionality not specific to diagramming, layer 2 provides the diagramming code to create and layout elements and DSL constructs independent of a diagram type, and layer 3 provides diagram type-specific modules.
In the following, we describe selected features of our internal DSL.
We focus on (1)~defining a class with attributes and methods, (2)~defining an association between two classes, (3)~defining precise element layout, and (4)~styling elements, which are depicted in \Cref{fig:dsl-example}.
Control-flow structures like \texttt{for} loops can be used to efficiently create a layout for multiple elements grouped in a list. 
Class diagrams are specified within the \texttt{classDiagram} function.

\textit{Defining a class with attributes and methods:} 
A class is defined by the function \texttt{class} taking the class name as argument.
In the background, this creates a variable of the same name, which can be used afterward to manipulate the class or add associations, but assigning the class to a custom variable is possible, too.
Within the class function, we can use \texttt{private}, \texttt{public}, and other functions to define the class members with respective visibility modifiers.
Within these functions, attributes, and methods are modeled in the UML class diagram's notation, omitting the visibility.

\textit{Defining an association:}
We support all types of UML class diagram associations, e.g., normal association, composition, or inheritance.
To define an association, we use the respective infix operator.
While \texttt{extends} is for inheritance, all other associations with their navigation direction and explicit non-navigability use arrow-like operators.
For example, for a uni-directional association from a class \texttt{Menu} to a class \texttt{Dish}, we can use the \verb|-->| operator in between both class variables.

\textit{Defining an element layout:}
The layout for different diagram elements is defined by different functions.
We manipulate the layout using the \texttt{layout} scope function of the \texttt{class} function for a class element.
Inside this function, we can define an absolute or relative position, width, and length of the rendered canvas.
The \texttt{with} scope function can be used to specify a path, segments, and labels for a canvas connection, i.e., association in the class diagram context.
Labels could be multiplicities and are positioned relative to the canvas connection.
We support line segments, axis-aligned line segments, and bezier curves for the association layout. 

\textit{Defining an element style:}
The \texttt{styles} scope function can be applied to any element to define element-specific styles, e.g., font size or fill color, in a declarative way.

\begin{figure}
    \centering
    \includegraphics[width=0.95\linewidth]{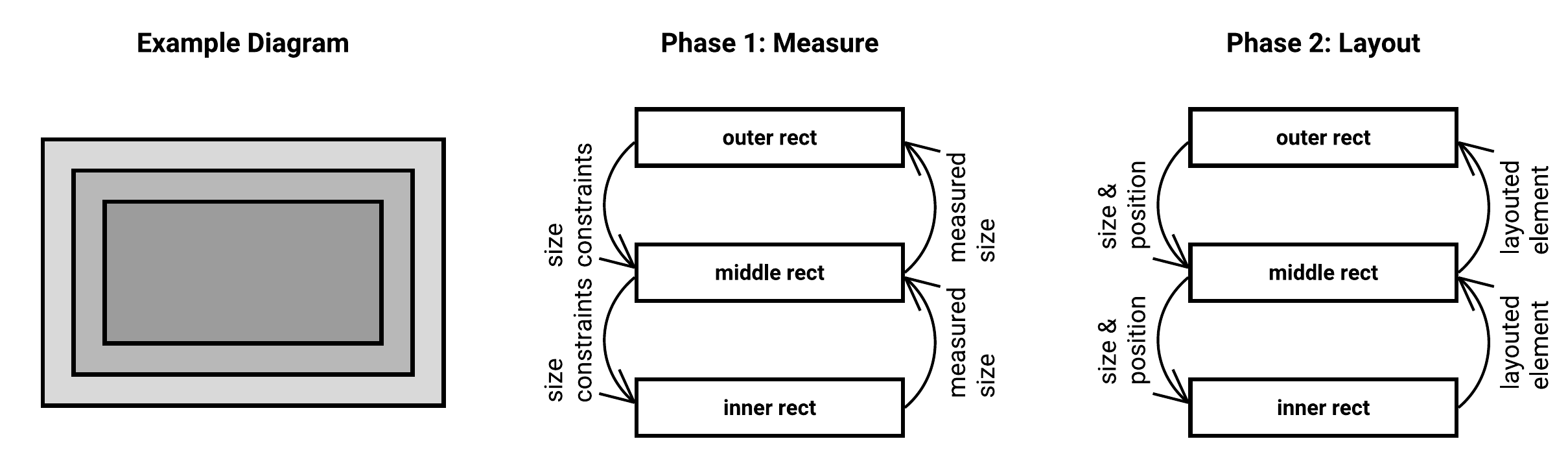}
    \caption{Visualization of the layout algorithm used.}
    \label{fig:layout-process}
\end{figure}

To be used for diagramming, the diagram defined using our DSL needs to be rendered to a visual result.
Overall, the rendering pipeline works as follows:
First, the diagram code is parsed and executed.
The result is a HyLiMo diagram defined using the data structures provided by the Layer 2 diagram module.
Conceptually, the diagram is a triple of a tree of diagram elements, a list of style rules, and a set of font files.
Both the element tree and style rules take inspiration from web technologies, HTML and CSS, respectively.
Next, the layout process transforms the diagram into a so-called layouted diagram.
Layouted diagrams differ in three major ways from diagrams:
First, while diagrams use mainly relative positions, e.g., to stack elements vertically using a vbox, layouted diagrams primarily use absolute positioning and sizing.
Second, no separate style rules exist, instead only local attributes on elements are used, with layouting related attributes being omitted completely.
Third, layout-only elements are removed as they are no longer needed due to absolute positioning.
To create layouted diagrams, first, style rules are evaluated to obtain style attributes for all diagram elements.
Next, a two-phase layouting process inspired by Flutter and Windows Presentation Foundation (WPF) determines the size and position of retained elements, as shown in \Cref{fig:layout-process}.
Phase 1, called measure, passes size constraints down the element tree, with the calculated measuredSize traveling up the tree.
Phase 2, called layout, creates the layouted diagram.
Here, size and position travel down the tree, while the resulting layouted elements travel up the tree.
Last, the layouted diagram serves as an intermediary format for all target platforms.
Using a visitor-based approach, it can be transformed into either an SVG or PDF file or displayed in the interactive graphical view part of the hybrid editor.
A noteworthy implementation detail of our approach is that the whole rendering pipeline neither depends on browser-only nor any native dependencies, thus producing consistent results in both a browser and Node.js environment.

\section{IDE Integration}\label{sec:ide-integration}
\begin{figure}
    \centering
    \includegraphics[width=0.95\linewidth]{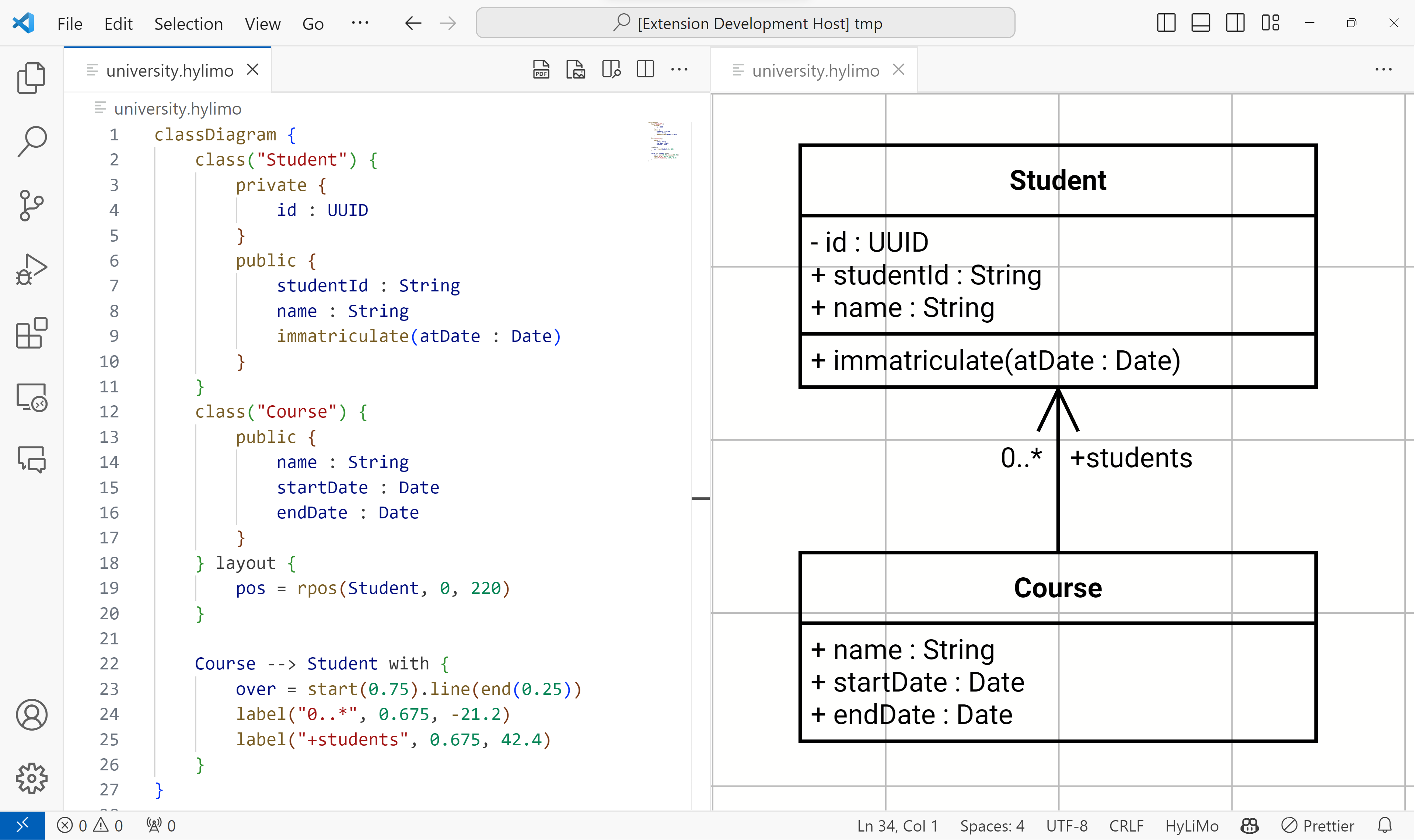}
    \caption{Example diagram in the HyLiMo VS Code extension.}
    \label{fig:ide-extension}
\end{figure}

\begin{figure}
    \centering
    \includegraphics[width=0.95\linewidth]{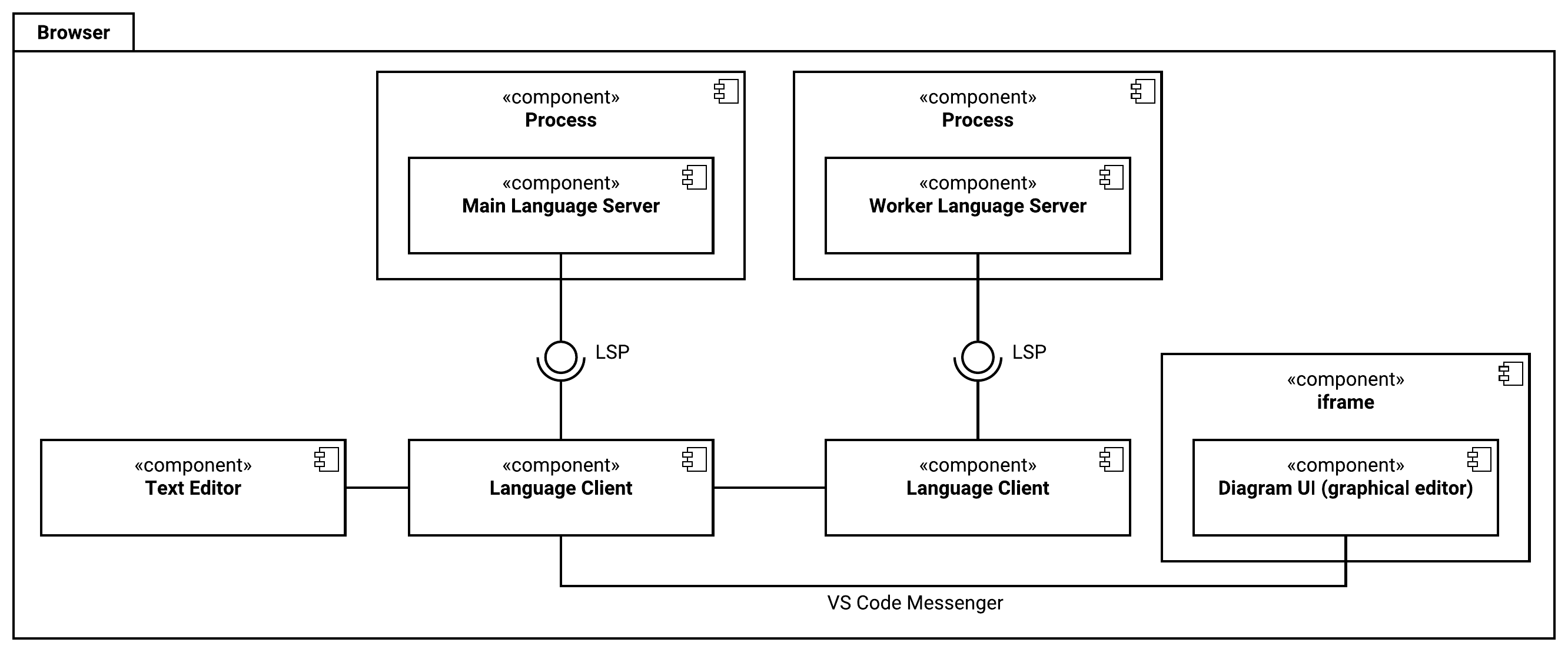}
    \caption{Architecture of our HyLiMo VS Code extension.}
    \label{fig:architecture}
\end{figure}

In addition to a regular web editor, we implemented HyLiMo\footnote{\url{https://github.com/hylimo/hylimo} and \url{https://github.com/hylimo/hylimo-vsc}} as an IDE extension for VS Code, which we include in VS Code's extension marketplace\footnote{\url{https://marketplace.visualstudio.com/items?itemName=NiklasKrieger.hylimo-vsc}}.
\Cref{fig:ide-extension} depicts an example diagram in the VS Code extension.
Left is the textual editor, and right is the interactive graphical editor.
We offer explicit functionality for exporting the rendered diagram in multiple formats, syntax highlighting, auto-formatting, and auto-completion.
Like the web version, it also supports diagnostics (syntactical and runtime errors) and reveals the source code when double-clicking a graphical element.

Our VS Code extension's architecture consists of the \textit{Text Editor}, \textit{Diagram UI} (graphical editor), and currently, two language clients connected to language servers via LSP, as shown in \Cref{fig:architecture}.
As mentioned, the language servers work in a main/worker architecture where the main language server is responsible for lightweight tasks.
All diagram executions are proxied to the worker language server by the main language server.
The main language server's language client is responsible for all diagram files currently opened in the \textit{Text Editor} and communicates via the \textit{VS Code Messenger} with the \textit{Diagram UI}, which runs isolated in an iframe.

\section{Evaluation}\label{sec:evaluation}
We evaluated our DSL and hybrid diagramming editor based on two use case studies in which one modeler each used HyLiMo to create UML class diagrams. 
The first use case's diagram is defined by 34 classes and three enumerations over 607 lines of textual definition.
The modeler particularly used lists and loops to equally space elements and transfer multiple connections into one at a specific point.
For the general layout creation, the modeler extensively used the graphical editor and fine-tuned it via the DSL.
The second use case consists of two smaller and one larger diagrams.
The modeler used features similar to the first use case's modeler.
We asked both modelers to share information on problems, feature requests, and general information about the framework's and editor's aspects, which worked well and did not work well.
In general, both modelers were satisfied with the diagramming experience in HyLiMo.
The detailed presentation and discussion of our evaluation can be found in~\cite{krieger2023hylimo}.
Please note that both modelers used the web version instead of the IDE integration, as the latter did not exist then.

While our evaluation provides some basic validation indicating that a hybrid diagramming approach, including style and layout information in the DSL, enables fast and precise diagramming, the results might not be generalizable due to the limited sample size.
Therefore, we require further, more in-depth evaluation.
In a controlled study, we plan to include more participants and measure diagramming time with HyLiMo and other editors.
We further plan to evaluate the IDE extension in a qualitative user survey.

\section{Related Work}\label{sec:related-work}
In this section, we discuss related work.
We start by elaborating on related research approaches, then introduce modeling and diagramming tools focusing on IDE integration and blended modeling.

Ciccozzi et al.~\cite{blended1} present the concept of blended modeling to improve the modeling experience by offering multiple notations simultaneously to the modeler. 
The concept is picked up by David et al.~\cite{blended2}, who conducted a systematic study on blended modeling and identified that only 62\% of the editors synchronize the different notations.
Addazi et al.~\cite{blendedprofile} propose a framework for blended textual and graphical modeling based on the \textit{Eclipse Modeling Framework} (EMF).
In their evaluation, they show that, depending on the task, either textual or graphical notation is more efficient and, therefore, hybrid modeling leads to faster modeling. 
Also, Cooper et al.~\cite{hybridsirius} outline an overview of hybrid modeling in \textit{EMF} and identify multiple requirements for it.
Alternative to \textit{EMF}-based approaches, Glaser et al.~\cite{glaser2021biger} present a hybrid modeling approach for entity-relationship diagrams, which they offer as Visual Studio Code extension.
Like our approach, they use the \textit{LSP} for their server and \textit{Eclipse Sprotty} for their graphical view.
Walsh et al.~\cite{walsh1} present an architecture for hybrid modeling languages using \textit{LSP} and the \textit{Graphical Language Server Platform}.
Similar to Addazi et al.~\cite{blendedprofile}, they propose a shared underlying model that is bidirectionally mapped to the textual and graphical view.
While all those works provide a blended modeling or diagramming concept, none allows the engineer to create generic and arbitrary diagrams.
Instead, they are restricted to specific use cases.
Furthermore, none of the works supports general-purpose programming language elements like loops for more convenient diagramming, nor do they store style and layout information in the textual notation.

There are various tools for modeling or diagramming.
We identified the general tools primarily based on the review by Ozkaya~\cite{umltools} and grey literature.
While more tools exist, this paper focuses on editors with an IDE extension or approaches similar to our editor's.
In \textit{Mermaid} and \textit{PlantUML}, the modeler creates specific diagrams, e.g., UML diagrams, using a textual syntax, which are then rendered by the editor.
\textit{Mermaid} offers a Visual Studio Code and IntelliJ extension\footnote{Mermaid: \url{https://bit.ly/mermaid-vs-code} and \url{https://bit.ly/mermaid-intellij}}.
While in the VS Code extension, modelers create the mermaid diagrams in markdown files, and the extension renders it, the IntelliJ extension additionally offers syntax highlighting and special file-endings.
For Mermaid, also other VS Code extensions exist.
To model \textit{PlantUML}, there are extensions for VS Code, IntelliJ, and Eclipse\footnote{PlantUML: \url{https://bit.ly/plantuml-vscode}, \url{https://bit.ly/plantuml-intellij}, and \url{https://bit.ly/plantuml-eclipse}}.
They differ in features but generally offer a preview for the rendered diagram, syntax highlighting, export, and other quality-of-life features.
Specific to Eclipse, the extension allows diagram rendering for Java classes, Ecore models, and OSGi manifests. 
\textit{Mermaid} and \textit{PlantUML} render the diagrams with a strict auto-layout which can be influenced using invisible edges, but cannot be manually changed.
While both offer IDE integrations to create diagrams more efficiently, none applies a blended modeling or diagramming approach.

Instead of creating the diagrams via a textual syntax, in \textit{draw.io}, the modeler uses a graphical editor to add and edit shapes.
It allows more general diagrams than Mermaid and \textit{PlantUML}, and layout is done manually, but modeling takes more time, and keeping a precise layout over time when the diagram evolves is cumbersome and time-consuming.  
In addition to an IntelliJ plugin, \textit{draw.io} offers an integration to VS Code\footnote{draw.io: \url{https://bit.ly/drawio-vscode}} too, which uses the offline version of the regular \textit{draw.io} editor and especially allows code linking and collaborative editing. 
It further enables editing the XML itself in a hybrid approach.
\textit{UMLet}~\cite{umlet} is a hybrid diagramming editor, available as standalone version, Eclipse plugin, and recently also as web-based version and VS Code plugin.
It combines textual and graphical notations to model UML diagrams, with both notations focusing on different aspects of the diagram.
While both editors allow manual layouting, they do not support the user with creating the layout efficiently.
Furthermore, they do not offer a full hybrid modeling/diagramming approach. 

In \textit{Diagrams}, the modeler creates system architecture diagrams using Python as an internal DSL.
A VS Code plugin providing diagram preview is available\footnote{Diagrams: \url{https://bit.ly/diagrams-vscode}}.
Regarding diagramming tools without IDE integration but with relevant aspects, we identified  \textit{TikZ-UML}.
In \textit{TikZ-UML}, the modeler specifies layout information manually via \textit{TikZ}'s textual syntax.
However, the modeler cannot move and layout the diagram via a rendered version, which makes creating a precise layout time-consuming.

\section{Conclusion}\label{sec:conclusion}
In this paper, we propose a flexible hybrid textual and graphical diagramming editor for efficiently creating diagrams such as UML class diagrams when manual control over the layout is required to communicate information efficiently.
In our tool, the textual and graphical editors are live synchronized, and the textual syntax stores layout and styling information as code.
We briefly presented the DSL for the concrete textual syntax, the underlying general-purpose programming language, and our hybrid editor.
While our DSL allows the creation of arbitrary diagrams, we offer specific features for straightforward UML class diagram diagramming.
We integrated the editor into the Visual Studio Code IDE, enabling developers to create diagrams more easily and allowing for version control.
A user study indicates that using such hybrid diagramming results in more efficient and effective diagramming whenever a precise layout is required.
Nevertheless, we plan to conduct a more in-depth study to validate our approach in future work.
Additionally, we intend to extend the DSL for other diagram types, e.g., UML component and sequence diagrams.

\bibliographystyle{ACM-Reference-Format}
\bibliography{bibliography}


\end{document}